\documentclass[12pt]{article}
\usepackage{amssymb}

\input{tcilatex}
\begin{document}

\title{\textbf{Locally Rotationally Symmetric Vacuum Solutions in f(R) Gravity}}
\author{ M. Jamil Amir \thanks{%
mjamil.dgk@gmail.com} and Sadia Sattar \thanks{%
sadiasattar30@yahoo.com} \\
Department of Mathematics, University of Sargodha,\\
Sargodha-40100, Pakistan.}
\date{}
\maketitle

\begin{abstract}
This paper is devoted to find the Locally Rotationally Symmetric
(LRS) vacuum solutions in the context of $f(R)$ theory of
gravity. Actually, we have considered the three metrics
representing the whole family of LRS spacetimes and solved the
field equations by using metric approach as well as the assumption
of constant scalar curvature. It is mention here that $R$ may be
zero or non-zero. In all we found $10$ different solutions.
\end{abstract}


\section{Introduction}

The rapid growth of the universe is one of the undo crisis in the
cosmology \cite{1sp1}. Some of researchers consider this rushing
growth of universe is may be due to some unknown energy momentum
component. The equation of state of energy momentum component is
$P=\omega \rho $. A number of theoretical models such as
quintessential scenarios \cite{2sp3}, which generalize the
cosmological constant approach \cite{3sp4}, higher dimensional
scenarios \cite{4sp5},\cite{5sp6} or alternative to cosmological
fluids with exotic equation of state \cite{6sp7} have been given
to solve a problem which untilled seems to be still unresolved.

Since General Relativity (GR) fails to explain the acceleration of
the universe. Thus, we are compelled to introduced some kind of
dark matter (DM) or dark energy (DE), which are responsible for
the rapid growth of the universe \cite{7in25}. DE and DM models
have been inspected in relative to their ability of explaining the
acceleration of the universe, however until now there are no
experimental indications for the existence of the predicted amount
of dark energy in the universe.

So, there is a need of modified or generalized theories to solve
the above mention problem. One of these modified theories that
arose a lot of stimulation is called $f(R)$ theory of gravity
which is obtained by simply replacing $R$ with the general
function $f(R)$ in the Einstein-Hilbert Lagrangian of GR. The
resultant field equations become more complicated as well as of
higher order due to the use of general function $f(R)$ in the
action. Thus, more exact solutions are expected in this theory
than GR, due to higher order derivative. The Ricci scalar $R$ and
trace of the energy-momentum tensor $T$ have a differential
relationship in this theory whereas, in GR, they are algebraically
related, i.e., $R=-\kappa T$. Moreover, in this theory Birkhoff's
theorem does not hold \cite{8t3} \ and also $T=0$ not imply $R=0$
as it is true in GR.

Weyl \cite{9f13} and Eddington \cite{10f14} were the first who
studied the  action in the context of $ f(R)$ theory of gravity.
Jakubiec and Kijowski \cite{11f16} investigated theories of
gravitation with non-linear Lagrangian. There is quite enough
material in literature in which  different issues have been
explored in $f(R)$ theories of gravity. Multamaki and Vilja
\cite{12f110} worked on spherically symmetric vacuum solutions in
$f(R)$ theory. The same authors \cite{13f111} also studied the
perfect fluid solutions and they found that pressure and density
did not uniquely determine the function $f(R)$.

In $2008$, the static cylindrically symmetric vacuum solutions in
metric $f(R)$ theory of gravity have been investigated by Azadi
and his coworkers \cite{14f114}. Momeni \cite{15f115} extended
this work to the non-static cylindrically symmetric solutions.
Sharif and Rizwana \cite{16fn12} explored the non-vacuum solution
of Bianchi type $VI_0$ universe in $f(R)$ gravity. Capozziello et
al. \cite{17fn13} investigated spherically symmetric solutions of
$f(R)$ theories of gravity by the Noether symmetry approach.
Reoboucas and Santos \cite{18fn17} analyzed G$\ddot{o}$del-type
universes in $f(R)$ gravity. Sharif and Shamir \cite{19fn18}
studied static plane symmetric vacuum solutions in $f(R)$ gravity.

Many authors studied Bianchi type spacetimes in different
frameworks. Kumar and Sing \cite{20fn28} investigated solutions of
the field equations in presence of perfect fluid using Bianchi
type $I$ spacetime in GR. Lorenz-Petzold \cite{21fn20} explored
exact Bianchi type $III$ solutions in the presence of the
electromagnetic field. Xing-Xiang \cite{22fn30} investigated
Bianchi type $III$ string cosmology with bulk viscosity. Wang
\cite{23fn31} studied strings cosmological models with bulk
viscosity in Kantowski-sachs spacetime.

Shri and Singh \cite{23aB} explored the analytical solutions of
the Einstein-Maxwell field equations for cosmological models of
LRS Bianchi type-II, VIII and IX. Pant and Oli \cite{23bB}
investigated two fluid Bianchi type-II cosmological models. Singh
and Kumar \cite{23cB} studied the Bianchi type-II cosmological
model by means of constant deceleration parameter. The massive
cosmic string in extent of BII model was discussed by Belinchon
\cite{23dB}-\cite{23dB1}. Pradhan et. al. \cite{23eB}, Kumar
\cite{23fB} and Yadav et. al. \cite{23gB}-\cite{23gB1} inspected
the LRS BII cosmological models in existence of massive cosmic
string and varying cosmological constant.

The study of Bianchi type models in alternative or modified
theories of gravity is also an fascinating discussion. Kumar and
Sing studied perfect fluid solutions using Bianchi type $I$
spacetimes in scalar tensor theory. Sing and his coworkers
\cite{24fn35} investigated some Bianchi type $III$ cosmological
models in frame work of scalar tensor theory. Sharif and Shamir
\cite{25fn38}-\cite{26fn39} investigated the solutions of Bianchi
type $I$ and $V$ spacetimes in framework of $f(R)$ gravity. Paul
and collaborators \cite{27fn37} studied FRW cosmologies in $f(R)$
gravity. Recently, Jamil and Saima \cite{50} explored the spatially
homogeneous rotating solution in f (R) gravity and Its energy contents.

In this paper, we explore the vacuum solutions of the whole family
of LRS spacetimes in metric f(R) gravity. The field equations are
solved by taking the assumption of constant scalar curvature. The
paper is organized as follows: In section 2 a brief introduction
of the field equations in metric version of f(R) gravity is given.
Section 3 contains LRS solutions in $f(R)$ theory of gravity,
especially, solutions with constant scalar curvature. In the last
section, we summarize the results.

\section{Field Equations in Metric $f(R)$ Gravity}

The three main approaches in f(R) theory of gravity are "Metric
Approach", "Palatini formalism" and " affine $f(R)$ gravity". In
metric approach, the connection is the Levi-Civita connection and
variation of the action is done with respect to the metric tensor.
While, in Palatini formalism, the metric and the connection are
independent of each other and variation is done for the two
mentioned parameters independently. In metric-affine $f(R)$
gravity, both the metric tensor and connection are treating
independently and assuming the matter action depends on the
connection as well. In our work, we use the metric approach only.
The action for $f(R)$ theory of gravity is given by
\begin{equation}\
S=\int \sqrt{-g}(\frac{1}{16\pi G}f(R)+L_{m})d^{4}x  \label{1}
\end{equation}
Here, $f(R)$ is the general function of the Ricci scalar and
$L_{m}$ is the matter Lagrangian. To derive the field equations
for the metric f(R) gravity, we vary the action given by
Eq.(\ref{1}) with respect to the metric tensor $g_{\mu \nu }\ $
and arrive at
\begin{equation}
F(R)R_{\mu \nu }-\frac{1}{2}f(R)g_{\mu \nu }-\nabla _{\mu }\nabla
_{\nu }F(R)+g_{\mu \nu }\Box F(R)=\kappa T_{\mu \nu }, \label{2}
\end{equation}
where $F(R)\equiv \frac{df(R)}{dR}$ and $\box \equiv \nabla ^{\mu
}\nabla _{\mu }$. Here, $\nabla _{\mu }$ represents the covariant
derivative and $T_{\mu \nu }$ is the standard matter
energy-momentum tensor which is derived from the Lagrangian
$L_{m}$. These are fourth order partial differential equations in
the metric tensor due to the last two terms on the left hand side
of the equation. These equations can be reduced to the field
equations of GR, i.e., EFEs by replacing $f(R)$ with $R$. To
contract the field equations (\ref{2}), we multiply it by the
suitable components of the inverse metric functions and obtain
\begin{equation}\label{3}
F(R)R-2f(R)+3 \Box F(R)=\kappa T.
\end{equation}
In vacuum, i.e., for $T=0$, the last equation takes the form as
\begin{equation}
F(R)R-2f(R)+3 \Box F(R)=0.  \label{4}
\end{equation}
This is an important equation as it will be helpful to find the
general function $f(R)$ and also will be used in simplifying the
field equations. It is clear from Eq.(\ref{4}) that any metric
with constant curvature, say $R=R_{0},$ is a solution of this
equation if the following condition holds
\begin{equation}
F(R_{0})R_{0}-2f(R_{0})=0.  \label{5}
\end{equation}
This is known as constant scalar curvature condition in vacuum.
While, for non-vacuum case, the constant scalar curvature
condition can be obtained from Eq.(\ref{3}) and is given by
\begin{equation}
F(R_{0})R_{0}-2f(R_{0})=\kappa T.  \label{6}
\end{equation}
The conditions (\ref{5}) and (\ref{6}) are very important to check
the stability conditions of the $f(R)$ models. If we differentiate
the Eq.(\ref{3}) w.r.t. $R$, we obtain
\begin{equation}
F^{\prime }(R)R-R^{\prime }F(R)+3(\Box F(R))^{\prime }=0,
\label{7}
\end{equation}
where prime represents derivative with respect to $R$. This
equation gives the consistency relation for $F(R)$.

\section{Locally Rotationally Symmetric Solutions in f(R) Gravity}

The LRS spacetimes which contain well-known exact solutions of the
EFEs are widely studied by many authors \cite{54i27}-\cite{56}. It
has been shown that they admit a group of motions $G_{4}$ acting
multiply transitively on three dimensional non-null orbits
space-like ($S_{3}$) or time-like ($T_{3}$) and the isotropy group
is a spatial rotation. The whole family of LRS spacetimes is
represented by the following three metrics:
\begin{eqnarray}
ds^{2} &=&\epsilon \lbrack 
-dt^{2}+A^{2}(t)dx^{2}]-B^{2}(t)dy^{2}-B^{2}(t)\Sigma
^{2}(y,k)dz^{2},
\label{8} \\
ds^{2} &=&\epsilon \lbrack -dt^{2}+A^{2}(t)\{dx-\Lambda
(y,k)dz\}^{2}]-B^{2}(t)dy^{2}  \nonumber \\
&-&B^{2}(t)\Sigma ^{2}(y,k)dz^{2},  \label{9} \\
ds^{2} &=&\epsilon \lbrack
-dt^{2}+A^{2}(t)dx^{2}]-e^{2x}B^{2}(t)(dy^{2}+dz^{2}), \label{10}
\end{eqnarray}
where $k=-1,0,1,~\epsilon =\pm 1$. Here $\Sigma$ and $\Lambda$ are
the multivalued functions depending upon the value of $k$ and are
defined as
\[
\Sigma =\left\{
\begin{array}{l}
\sin y,\quad ~~k=1, \\
y,\quad \quad \quad k=0, \\
\sinh y,\quad k=-1, \\
\end{array}
\right.
\]
and
\[
\Lambda =\left\{
\begin{array}{l}
\cos y,\quad ~~k=1, \\
\frac{y^{2}}{2},\quad \quad ~~k=0, \\
\cosh y,\quad k=-1. \\
\end{array}
\right.
\]
Further, it is mentioned here that corresponding to $\epsilon =1$
and $ \epsilon =-1$ we obtain the static and non-static LRS
solutions respectively. In this paper, we will discuss only the
non-static case as the results for the static case can be obtained
consequently. For $ \epsilon =-1$, the above equations take the
form
\begin{eqnarray*}
ds^{2} &=&dt^{2}-A^{2}(t)dx^{2}-B^{2}(t)dy^{2}-B^{2}(t)\Sigma
^{2}(y,k)dz^{2}, ~~~~~~ (\texttt{Metric-I})   \label{11} \\
ds^{2}
&=&dt^{2}-A^{2}(t)dx^{2}-B^{2}(t)e^{2x}dy^{2}-B^{2}(t)e^{2x}dz^{2},~~~~~~~~
(\texttt{Metric-II})\label{12}\\
ds^{2} &=&dt^{2}-A^{2}(t)dx^{2}-B^{2}(t)dy^{2}-\{A^{2}(t)\Lambda
^{2}(y,k)
\nonumber \\
&+&B^{2}(t)\Sigma ^{2}(y,k)\}dz^{2}+2A^{2}(t)\Lambda
(y,k)dxdz.~~~~~~~~~~~~~~~ (\texttt{Metric-III}) \label{13}
\end{eqnarray*}
It has been shown that, for $k=0$, the metric-I may reduce to
Bianchi types $I(BI)$ or $VII_{0}, (BVII_{0})$, for $k=-1$,
Bianchi type $III, (BIII)$ and Kantowski-Sachs (KS) for $k=+1$.
The metric-II represents Bianchi type $V(BV)$ or $VII_{h},
(BVII_{h})$ metric while the metric-III reduces to Bianchi types
$II(BII)$ for $k=0$, $VIII(BVIII)$ or $III(BIII)$ for $k=-1$ and
$IX(BIX)$ for $k=+1$.

\subsection{Solution of the Metric-I}

Now, we shall solve the metric-I by considering the following
three cases:\\

\textbf{a).} When $k=0\quad\quad\quad $\textbf{b).} When
$k=+1\quad\quad\quad $\textbf{c).} When $k=-1$

\subsubsection{Case a:}

For $k=0$, the metric-I takes the form
\begin{equation}
ds^{2}=dt^{2}-A^{2}(t)dx^{2}-B^{2}(t)(d^{2}y+y^{2}d^{2}z).
\label{14}
\end{equation}
After evaluating the components of Ricci tensor, the Ricci scalar
turns out to be
\begin{equation}
R=-2\frac{\ddot{A}}{A}-4\frac{\ddot{B}}{B}-4\frac{\dot{A}\dot{B}}{AB}-2\frac{
\dot{B}^{2}}{B^{2}},  \label{15}
\end{equation}
where dot represents derivative with respect to time. Writing the
Eq.(\ref{4}) in the form
\begin{equation}
f(R)=\frac{3\Box F(R)+F(R)R}{2}.  \label{16}
\end{equation}
Using this value of $f(R)$ in the the field equations (\ref{2})
and setting $T_{\mu\nu}=0$ (for vacuum solutions), we have
\begin{equation}
\frac{F(R)R_{\mu \nu }-\nabla _{\mu }\nabla _{\nu }F(R)}{ g_{\mu
\nu }}=\frac{F(R)R-\Box F(R)}{4}. \label{17}
\end{equation}
It is clear form equation (\ref{15}) that  the Ricci scalar
depends only on $t$ and hence $F(R)$ will be a function of $t$
only, that is, $F\equiv F(t)$. So, Eq.(\ref{17}) becomes the set
of ordinary differential equations involving $F(t),A(t)$ and
$B(t)$. Now, one can straightforward write from the L.H.S. of
Eq.(\ref {17}) that
\begin{equation}
E_{\mu }\equiv \frac{F(R)R_{\mu \mu }-\nabla _{\mu }\nabla _{\mu
}F(R)}{ g_{\mu \mu }}  \label{18},
\end{equation}
which is independent of the index $\mu $ and hence $E_{\mu
}-E_{\nu }=0$ for all $\mu $ and $\nu$. Thus, $E_{0}-E_{1}=0$
gives
\begin{equation}
-\frac{2\ddot{B}}{B}+\frac{2\dot{A}\dot{B}}{AB}+\frac{\dot{A}\dot{F}}{AF}-
\frac{\ddot{F}}{F}=0.  \label{19}
\end{equation}
Similarly, $E_{0}-E_{2}=0$ yields
\begin{equation}
-\frac{\ddot{A}}{A}-\frac{\ddot{B}}{B}+{\frac{\dot{B}^{2}}{B^{2}}}+\frac{
\dot{A}\dot{B}}{AB}+\frac{\dot{B}\dot{F}}{BF}-\frac{\ddot{F}}{F}=0
\label{20}
\end{equation}
while the remaining cases, obtained by varying $\mu $ and $\nu$,
yield the equations which are scalar multiple of the last two
equations. As we obtain two non-linear differential equations and
three unknowns, i.e., $A,~B$ and $F$ so we have to impose an other
condition to find the solution of these equations. We shall use
assumption of constant curvature and find the solution
of these equations as follows:\\
\textbf{Constant Curvature Solution}

For constant curvature solution, i.e., $R=R_{0},$ we have
\begin{equation}
\dot{F}(R_{0})=\ddot{F}(R_{0})=0. \label{21}
\end{equation}
Making use of this condition of constant curvature, the
Eqs.(\ref{19}) and (\ref{20}) reduce to
\begin{equation}
\frac{\ddot{B}}{B}-\frac{\dot{A}\dot{B}}{AB}=0,  \label{22}
\end{equation}
and
\begin{equation}
\frac{\ddot{A}}{A}+\frac{\ddot{B}}{B}-{\frac{\dot{B}}{B^{2}}}^{2}-\frac{\dot{
A}\dot{B}}{AB}=0  \label{23}
\end{equation}
respectively. Further, we will solve the last two equations by
using following two assumptions:

\quad \textbf{I.} \textit{Power law assumption} \quad\quad\quad
\textbf{II.} \textit{Exponential assumption}\\

\textbf{Case aI:}

Here we assume that $A\propto t^{m}$ and $B\propto t^{n}$, where
$m$ and $n$ are any real numbers. In other words, we substitute
$A=K_{1}t^{m}$ and $B=K_{2}t^{n}$ in Eqs.(\ref{22}) and
(\ref{23}), where $K_{1}$ and $K_{2}$ are constants of
proportionality, to obtain
\begin{equation}
n(n-1-m)=0,  \label{24}
\end{equation}
\begin{equation}
m^{2}-m-n-mn=0.  \label{25}
\end{equation}
The simultaneous solution of these equations gives two
possibilities for the values of $m$ and $n$
$$
i).~ m=\frac{-1}{3},~ n=\frac{2}{3}  \quad\quad\quad ii).~ m=1,~
n=0
$$
\textbf{Case aI(i):}

In this case, we have $A=K_1 t^{-\frac{1}{3}}$ and $B=K_2
t^{\frac{2}{3}}$. Then the corresponding solution turns out to be
\begin{equation}
ds^{2}=dt^{2}-(K_{1})^{2}t^{-\frac{2}{3}}dx^{2}-(K_{2})^{2}t^{\frac{4}{3}
}(dy^{2}+y^{2}dz^{2}). \label{26}
\end{equation}
It is mentioned here that the Ricci scalar becomes zero for this
case. This solution corresponds to the Kinematics self-similar
solution of the second kind for the tilted dust case given in
table $1$ of the \cite{56}.
\\
\textbf{case aI(ii):}

In this case, we obtain $A=K_1 t$ and $B=K_2 $ and arrive at the
solution
\begin{equation}
ds^{2}=dt^{2}-(K_{1})^{2}t^{2}dx^{2}-(K_{2})^{2}(dy^{2}+y^{2}dz^{2}).
\label{27}
\end{equation}
In this case, again the Ricci scalar vanishes. The solution given
in Eq.(\ref{27}) coincides to the Kinematics self-similar solution
of the first kind for the tilted dust case given in table $1$ of
the \cite{56}.

\textbf{Case aII:}

In this case, we assume that $A(t)=e^{2\mu (t)}$ and $%
B(t)=e^{2\lambda (t)}$ so that metric (\ref{14}) becomes
\begin{equation}
ds^{2}=dt^{2}-e^{4\mu (t)}dx^{2}-e^{4\lambda
(t)}[dy^{2}+y^{2}dz^{2}]. \label{28}
\end{equation}%
The corresponding Ricci scalar turns out to be
\begin{equation}
R=-8\dot{\mu}^{2}-4\ddot{\mu}-24\dot{\lambda}^{2}-8\ddot{\lambda}-16\dot{\mu}
\dot{\lambda}.  \label{29}
\end{equation}
Now, by substituting $A(t)=e^{2\mu (t)}$ and $ B(t)=e^{2\lambda
(t)}$ in Eqs.(\ref{22}) and (\ref{23}), we have
\begin{equation}
\ddot{\lambda}+2(\dot{\lambda )^{2}}-2\dot{\mu}\dot{\lambda} =0
\label{30} \\
\end{equation}
and
\begin{equation}
\ddot{\mu}+\ddot{\lambda}+2\dot{\mu}^{2}-2\dot{\mu}\dot{\lambda}
=0 \label{31}
\end{equation}
respectively. Eq.($\ref{30}$ ) can be written as
\begin{equation}
\dot{\lambda}(\frac{\ddot{\lambda}}{\dot{\lambda}}+2\dot{\lambda}-2\dot{\mu}
)=0,  \label{32}
\end{equation}
which yields the following two cases:
$$~~~~~~~~~~\texttt{i)}\ \dot{\lambda}=0,~~~~~~~~~~~~~~~\texttt{ii)}\
\frac{\ddot{\lambda}}{\dot{\lambda}}+2\dot{\lambda}-2
\dot{\mu}=0$$ We solve the Eqs.(\ref{22}) and (\ref{23}) ) for
these two cases as:\\
\textbf{Case aII(i):}

This case implies that
\begin{equation}
\lambda =a,  \label{33}
\end{equation}%
where a is constant of integration. By putting this value of
$\lambda$ in Eq.( \ref{31}), we obtain
\begin{equation}
\ddot{\mu}+2\dot{\mu}^{2}=0.  \label{34}
\end{equation}
On integration, the last equation implies that
\begin{equation}
\mu =ln(b\sqrt{t-c}),  \label{35}
\end{equation}%
where $b$ and $c$ are integration constants. Thus, the metric
(\ref{28}) takes the form
\begin{equation}
ds^{2}=dt^{2}-(b^{2}t-b^{2}c)^{2}dx^{2}-e^{4a}(dy^{2}+y^{2}dz^{2}).
\label{36}
\end{equation}\\
\textbf{Case aII(ii):}

In this case, the constraint equation
\begin{equation}
\frac{\ddot{\lambda}}{\dot{\lambda}}+2\dot{\lambda}-2\dot{\mu}=0
\label{37}
\end{equation}
yields the value of $\mu$ as
\begin{equation} \mu =\lambda
+\frac{1}{2}ln\dot{\lambda}+d.  \label{38}
\end{equation}
Substituting this value in Eq.(\ref{29}), we get the scalar
curvature as
\begin{equation}
R=-2\frac{\lambda
^{...}}{\dot{\lambda}}-28\ddot{\lambda}-48\dot{\lambda}^{2}.
\label{39}
\end{equation}
The last equation implies that, for constant scalar curvature, we
have to assume
\begin{equation}
-2\frac{\lambda
^{...}}{\dot{\lambda}}-28\ddot{\lambda}-48\dot{\lambda}
^ {2}=constant.  \label{40}
\end{equation}
This is a third order linear ordinary differential equation which
can not be analytically solved easily. So, we assume that $\lambda
(t)$ is a linear function of $t$, i.e., $ \lambda (t)=ft+g$, where
$f$ and $g$ are any arbitrary constants. Then Eq.(\ref{38}) gives
$\mu=ft+\bar{g}$, where $\bar{g}=g+\frac{1}{2}lnf+d$.
Consequently, the metric (\ref{28}) becomes
\begin{equation}
ds^{2}=dt^{2}-e^{4(ft+\bar{g})}dx^{2}-e^{4(ft+g)}[dy^{2}+y^{2}dz^{2}].
\label{41}
\end{equation}
The corresponding Ricci scalar takes the form
\begin{equation}
R=-48f^{2}.  \label{42}
\end{equation}
We can write the metric (\ref{41}) as
\begin{equation}
ds^{2}=dt^{2}-a^{2}e^{2t}dx^{2}-b^{2}e^{2t}[dy^{2}+y^{2}dz^{2}],
\label{42}
\end{equation}
where $a=e^{\bar{g}}/2f$ and $b=e^{g/2f}$. Hence, this solution
corresponds to the Kinematics self-similar solution of the zeroth
kind for the parallel vector field case \cite{56}. It is mentioned
here that the trivial solution of the Eq.(\ref{14}) coincides with
the Kinematics self-similar solution of the infinite kind for the
parallel dust case given in Eq.(39) of \cite{56}.

\subsubsection{Case b:}

For $k=1$, the Metric-I takes the form
\begin{equation}
ds^{2}=dt^{2}-A^{2}(t)dx^{2}-B^{2}(t)[dy^{2}+\sin ^{2}ydz^{2}].
\label{43}
\end{equation}
With the help of the components of Ricci tensor, the Ricci scalar
has been evaluated as
\begin{equation}
R=\frac{-2}{AB^{2}}[\ddot{A}B^{2}+2\ddot{B}AB+2\dot{A}\dot{B}B+A(\dot{B}
)^{2}+A].  \label{44}
\end{equation}
Eq.(\ref{18}) gives the following three independent equations for
the corresponding values of $\mu $ and $\nu$:

$E_{0}-E_{1}=0$ yields
\begin{equation}
-\frac{2\ddot{B}}{B}+\frac{2\dot{A}\dot{B}}{AB}+\frac{\dot{A}\dot{F}}{AF}-
\frac{\ddot{F}}{F}=0;  \label{45}
\end{equation}
$E_{0}-E_{2}=0$ gives
\begin{equation}
-\frac{\ddot{A}}{A}-\frac{\ddot{B}}{B}+\frac{(\dot{B})^{2}}{B^{2}}+\frac{
\dot{A}\dot{B}}{AB}+\frac{1}{B^{2}}+\frac{\dot{B}\dot{F}}{BF}-\frac{\ddot{F}
}{F}=0  \label{46}
\end{equation}
and $E_{1}-E_{2}=0$ implies that
\begin{equation}
-\frac{\ddot{A}}{A}+\frac{\ddot{B}}{B}-\frac{\dot{A}\dot{B}}{AB}+\frac{(\dot{
B})^2}{B^{2}}+\frac{1}{B^{2}}+\frac{\dot{B}\dot{F}}{BF}-\frac{\dot{A}\dot{F}}{
AF}=0.  \label{47}
\end{equation}
We will also use the constant curvature assumption to solve this
system of three equations\\
\textbf{Constant Curvature Solution:}

Using condition(\ref{21}), Eqs.(\ref{45})-(\ref{47}) reduce to
\begin{eqnarray}\label{48}
\frac{\ddot{B}}{B}-\frac{\dot{A}\dot{B}}{AB}&=&0;\\\label{49}
\frac{\ddot{A}}{A}+\frac{\ddot{B}}{B}-\frac{(\dot{B})^{2}}{B^{2}}-\frac{\dot{
A}\dot{B}}{AB}-\frac{1}{B^{2}}&=&0
\end{eqnarray}
and
\begin{equation}
-\frac{\ddot{A}}{A}+\frac{\ddot{B}}{B}-\frac{\dot{A}\dot{B}}{AB}+\frac{(\dot{
B})^2}{B^{2}}+\frac{1}{B^{2}}=0.  \label{50}
\end{equation}
We solve the equation (\ref{48}) by using power law assumption and
substitute $ A=K_{1}t^{m}$ and $B=K_{2}t^{n}$, where $K_{1}$ and
$K_{2}$ are constants of proportionality. Consequently, we have
\begin{equation}
n(m-n+1)=0.  \label{51}
\end{equation}
Here only  the case $n=0$ gives constant curvature so we leave the
case for which $m-n+1=0$. Substituting $n=0$, i.e., $B=K_2$ and
$A=K_{1}t^{m}$ in Eqs.( \ref{49}) and (\ref{50}), we obtain the
following single equation
\begin{equation}
m(m-1)-\frac{t^2}{K_2}=0.  \label{52}
\end{equation}
Comparing the co-efficient of $t^{0}$, we have $m(m-1)=0$, i.e.,
$m=0$ or $ m=1$, which  yields the only non-trivial solution when
$A=k_{1}t$ and $B=k_{2}$. Hence, the metric (\ref{43}) takes the
form
\begin{equation}
ds^{2}=dt^{2}-(K_{1}t)^{2}dx^{2}-(K_{2})^{2}[dy^{2}+\sin
^{2}ydz^{2}]. \label{53}
\end{equation}
The corresponding Ricci scalar is given by $R=-2/k_2^2$. The
trivial solution of the Eq.(\ref{43}) is exactly the same as the
Kinematics self-similar solution of the infinite kind given in
Eq.(32) of \cite{56}.

\subsubsection{Case c:}

Now, we discuss the case when $k=-1$, for which the metric-I takes
the form
\begin{equation}
ds^{2}=dt^{2}-A^{2}(t)dx^{2}-B^{2}(t)[dy^{2}+\sinh ^{2}ydz^{2}].
\label{54}
\end{equation}
With the help of the components of Ricci tensor, the Ricci scalar
becomes
\begin{equation}
R=\frac{-2}{AB^{2}}[\ddot{A}B^{2}+2\ddot{B}AB+2\dot{A}\dot{B}B+A(\dot{B}
)^{2}+A].  \label{55}
\end{equation}
Eq.(\ref{18}) gives the following three independent equations for
the corresponding values of $\mu $ and $\nu$:\\
$E_{0}-E_{1}=0$ yields
\begin{equation}
-\frac{2\ddot{B}}{B}+\frac{2\dot{A}\dot{B}}{AB}+\frac{\dot{A}\dot{F}}{AF}-
\frac{\ddot{F}}{F}=0;  \label{56}
\end{equation}
$E_{0}-E_{2}=0$ gives
\begin{equation}
-\frac{\ddot{A}}{A}-\frac{\ddot{B}}{B}+\frac{(\dot{B})^{2}}{B^{2}}+\frac{
\dot{A}\dot{B}}{AB}-\frac{1}{B^{2}}+\frac{\dot{B}\dot{F}}{BF}-\frac{\ddot{F}
}{F}=0  \label{57}
\end{equation}
and $E_{1}-E_{2}=0$ implies that
\begin{equation}
-\frac{\ddot{A}}{A}+\frac{\ddot{B}}{B}-\frac{\dot{A}\dot{B}}{AB}+\frac{(\dot{
B})^{2}}{B^{2}}-\frac{1}{B^{2}}+\frac{\dot{B}\dot{F}}{BF}-\frac{\dot{A}\dot{F
}}{AF}=0.  \label{58}
\end{equation}
Again we use the constant curvature assumption to solve these
equations
\\
\textbf{Constant Curvature Solution:}

Using the condition (\ref{21}), Eqs.(\ref{56})-(\ref{58}) reduce to
\begin{eqnarray}
\frac{\ddot{B}}{B}-\frac{\dot{A}\dot{B}}{AB}=0,  \label{59}\\
\frac{\ddot{A}}{A}+\frac{\ddot{B}}{B}-\frac{(\dot{B})^{2}}{B^{2}}-\frac{\dot{
A}\dot{B}}{AB}+\frac{1}{B^{2}}=0,  \label{60}\\
-\frac{\ddot{A}}{A}+\frac{\ddot{B}}{B}-\frac{\dot{A}\dot{B}}{AB}+\frac{(\dot{
B})^2}{B^{2}}-\frac{1}{B^{2}}=0.  \label{61}
\end{eqnarray}
We solve these equations by using the power law assumption and
substitute $A=K_{1}t^{m}$ and $B=K_{2}t^{n}$, where $K_{1}$ and
$K_{2}$ are constants of proportionality. Only the case when $n=0$
and $m=1$ gives constant curvature solution. Then, the metric
(\ref{54}) takes the form
\begin{equation}
ds^{2}=dt^{2}-(K_{1}t)^{2}dx^{2}-(K_{2})^{2}[dy^{2}+\sinh
^{2}ydz^{2}]. \label{62}
\end{equation}
In this case, the Ricci scalar is given by $R=2/k_2^2$.

The trivial solution of the Eq.(\ref{54}) is exactly the same as
the Kinematics self-similar solution of the infinite kind given in
Eq.(33) of \cite{56}.

\subsection{Solution of Metric-II}

With the help of the components of Ricci tensor, Ricci scalar for
the Metric-II turns out to be
\begin{equation}
R=-2[\frac{\ddot{A}}{A}+2\frac{\ddot{B}}{B}+\frac{(\dot{B})^{2}}{B^{2}}+2
\frac{\dot{A}\dot{B}}{AB}-\frac{3}{A^{2}}].  \label{63}
\end{equation}
From Eq.(\ref{18}), $E_{0}-E_{1}=0$ gives
\bigskip\
\begin{equation}
(-2\frac{\ddot{B}}{B}+2\frac{\dot{A}\dot{B}}{AB}-\frac{2}{A^{2}})F(R)+(\frac{
\dot{A}}{A}\dot{F}-\ddot{F})=0;  \label{64}
\end{equation}
$E_{0}-E_{2}=0$ yields
\bigskip
\begin{equation}
(\frac{\ddot{A}}{A}-\frac{\ddot{B}}{B}+\frac{(\dot{B})^{2}}{B^{2}}+\frac{
\dot{A}\dot{B}}{AB}-\frac{2}{A^{2}})F(R)+(\frac{\dot{B}}{B}\dot{F}-\ddot{F}
)=0  \label{65}
\end{equation}
and $E_{1}-E_{2}=0$ implies that
\begin{equation}
(-\frac{\ddot{A}}{A}+\frac{\ddot{B}}{B}+\frac{(\dot{B})^{2}}{B^{2}}-\frac{
\dot{A}\dot{B}}{AB})F(R)+(\frac{\dot{B}}{B}\dot{F}-\frac{\dot{A}}{A}\ddot{F}
)=0,  \label{66}
\end{equation}
which are the only independent equations for all possible cases.\\
\textbf{Constant Curvature Solution:}\\

Now, for constant curvature solution, the
Eqs.(\ref{64})-(\ref{66}) reduce to
\begin{eqnarray}
-2\frac{\ddot{B}}{B}+2\frac{\dot{A}\dot{B}}{AB}-\frac{2}{A^{2}}=0,
\label{67}\\
-\frac{\ddot{A}}{A}-\frac{\ddot{B}}{B}+\frac{(\dot{B})^{2}}{B^{2}}+\frac{\dot{
A}\dot{B}}{AB}-\frac{2}{A^{2}}=0,  \label{68}
\\
-\frac{\ddot{A}}{A}+\frac{\ddot{B}}{B}+\frac{(\dot{B})^{2}}{B^{2}}-\frac{\dot{
A}\dot{B}}{AB}=0.  \label{69}
\end{eqnarray}
By subtracting Eq.(\ref{68}) form Eq.(\ref{69}) we get the only
independent Eq.(\ref{67}). Using Eq.(\ref{67}) in Eq.(\ref{63}),
the corresponding Ricci scalar becomes
\begin{equation}
R=-2\frac{\ddot{A}}{A}-10\frac{\ddot{B}}{B}-2\frac{(\dot{B})^{2}}{B^{2}}+2
\frac{\dot{A}\dot{B}}{AB}.  \label{70}
\end{equation}
Now, we assume that $A(t)=e^{2a}$ and $B(t)=e^{2\lambda (t)}$ so
that Eq.(\ref{70}) takes the form
\begin{equation}
R=-28\ddot{\lambda}-48\dot{\lambda}^{2}.  \label{71}
\end{equation}
But according to our assumption, the scalar curvature must be
constant, i.e.,
\begin{equation}
cons\tan t=-28\ddot{\lambda}-48\dot{\lambda}^{2}  \label{72}
\end{equation}
which implies that $\lambda (t)$ must be a linear function of $t$,
i.e., $\lambda(t)=ft+g$, where $f$ and $g$ are any arbitrary
constants. Consequently, the metric takes the form
\begin{equation}
ds^{2}=dt^{2}-e^{4a}dx^{2}-e^{4(f\ t\ +\ g)}\ e^{2\
x}(dy^{2}+dz^{2}). \label{73}
\end{equation}
The corresponding Ricci scalar becomes $R=-24f^{2}+6e^{-4a}$.

It is mentioned here that, for the trivial case of the metric-II,
i.e., when $A(t)$ and $B(t)$ are both taken to be constant, the
solution corresponds to the Kinematics self-similar solution of
the infinite kind for the parallel vector field given in the
Eq.(50) of \cite{56}.

\subsection{Solution of Metric-III}

In this section, we will solve the Metric-III by considering the
following three cases:

~~~~$\mathbf{\alpha)}$\textbf{.} When $k=0\qquad \mathbf{\beta)
}$\textbf{.} When $k=+1\ \ \ \mathbf{\gamma) }$\textbf{.} When
$k=-1$
\subsubsection{Case $~\alpha$:} For $k=0$, the metric-III can
be written as
\begin{equation}
ds^{2}=dt^{2}-A^{2}(t)dx^{2}-B^{2}(t)dy^{2}-\left\{
A^{2}(t)\frac{y^{4}}{4} +B^{2}(t)y^{2}\right\}
dz^{2}+2A^{2}(t)dxdz.  \label{74}
\end{equation}
The corresponding Ricci scalar has been evaluated, by using the
components of the Ricci tensor, as
\begin{equation}
R=\frac{-2}{AB^{4}}[\ddot{A}B^{4}+2\ddot{B}AB^{3}+2\dot{A}\dot{B}B^{3}+A(
\dot{B})^{2}B^{2}-\frac{A^{3}}{4}].  \label{75}
\end{equation}%
The Eq.(\ref{18}) gives the following independent three equations
for the possibilities, $E_{0}-E_{1}=0$, $E_{0}-E_{2}=0$ and
$E_{1}-E_{2}=0$
\begin{eqnarray}
-\frac{2\ddot{B}}{B}+\frac{2\dot{A}\dot{B}}{AB}+\frac{A^{2}}{2B^{4}}+\frac{
\dot{A}\dot{F}}{AF}-\frac{\ddot{F}}{F}&=&0,  \label{76}\\
-\frac{\ddot{A}}{A}-\frac{\ddot{B}}{B}+\frac{(\dot{B})^{2}}{B^{2}}+\frac{
\dot{A}\dot{B}}{AB}-\frac{A^{2}}{2B^{4}}+\frac{\dot{B}\dot{F}}{BF}-\frac{
\ddot{F}}{F}&=&0,  \label{77}\\
-\frac{\ddot{A}}{A}+\frac{\ddot{B}}{B}-\frac{\dot{A}\dot{B}}{AB}+\frac{(\dot{
B})^{2}}{B^{2}}-\frac{A^{2}}{B^{4}}+\frac{\dot{B}\dot{F}}{BF}-\frac{\dot{A}
\dot{F}}{AF}&=&0  \label{78}
\end{eqnarray}
respectively.\\
\textbf{Constant curvature Solution:}

Using the condition of constant curvature given by equation
(\ref{21}) in the Eqs.(\ref{76})-(\ref{78}), we obtain

\begin{eqnarray}
\frac{-2\ddot{B}}{B}+\frac{2\dot{A}\dot{B}}{AB}+\frac{A^{2}}{2B^{4}}=0,\label{80}
\\
-\frac{\ddot{A}}{A}-\frac{\ddot{B}}{B}+\frac{(\dot{B})^{2}}{B^{2}}+\frac{
\dot{A}\dot{B}}{AB}-\frac{A^{2}}{2B^{4}}=0,  \label{81}
\\
-\frac{\ddot{A}}{A}+\frac{\ddot{B}}{B}-\frac{\dot{A}\dot{B}}{AB}+\frac{(\dot{
B})}{B^{2}}-\frac{A^{2}}{B^{4}}=0.  \label{82}
\end{eqnarray}
It is noticed that when we subtract Eq.(\ref{82}) from
Eq.(\ref{81}), it yields the Eq.(\ref{80}). Now, we solve
Eq.(\ref{80}) by assuming $A(t)=B^{2}(t)$ and arrive at
$$
\frac{\ddot{B}}{B}=\frac{1}{4}+\frac{2(\dot{B})^{2}}{B^{2}}.
$$
We further assume that $B(t)=e^{2\lambda (t)}$ so that the
corresponding Ricci scalar turns out to be
\begin{equation}
R=-120\dot{\lambda}^{2}-\frac{3}{2}.  \label{83}
\end{equation}
Now, to attain the assumption of scalar constant curvature, we
must put the following constraint
\begin{equation}
cons\tan t=-120\dot{\lambda}^{2}-\frac{3}{2},  \label{84}
\end{equation}
which implies that $\lambda(t)$ must be a linear function of $t$,
that is, $\lambda(t)=ft+g$, where $f$ and $g$ are any arbitrary
constants. Consequently, we get the following solution
\begin{eqnarray}
ds^{2}&=&dt^{2}-e^{8(ft+g)}dx^{2}-e^{4(ft+g)}dy^{2}\nonumber\\
&-&e^{4(f\ t\ +\ g)}\left\{\frac{y^4e^{4(ft+g)}}{4} +y^2\right\}
dz^{2} +2e^{8(ft+g)}dxdz. \label{85}
\end{eqnarray}
In this case, the corresponding Ricci scalar becomes
$R=-120f^{2}-\frac{3}{2}$.

\subsubsection{Case $~\beta$:}

In this case, i.e., for $k=1$, the Metric-III takes the form
\begin{eqnarray}
ds^{2}&=&dt^{2}-A^{2}(t)dx^{2}-B^{2}(t)dy^{2}-\left\{ A^{2}(t)\cos
^{2}y+B^{2}(t)\sin ^{2}y\right\} dz^{2}\nonumber\\
&+&2\cos yA^{2}(t)dxdz. \label{86}
\end{eqnarray}
The corresponding Ricci scalar, evaluated with the help of the
components of Ricci tensor, turns out to be
\begin{equation}
R=\frac{-1}{2AB^{4}}[4\ddot{A}B^{4}+8\ddot{B}AB^{3}+8\dot{A}\dot{B}B^{3}+4A(
\dot{B})^{2}B^{2}+4AB^{2}-A^{3}].  \label{87}
\end{equation}
The Eq.(\ref{18}) yields the only two independent equation, given
below:\\
 $E_{0}-E_{1}=0$ yields
\begin{equation}
-\frac{2\ddot{B}}{B}+\frac{2\dot{A}\dot{B}}{AB}+\frac{A^{2}}{2B^{4}}+\frac{
\dot{A}\dot{F}}{AF}-\frac{\ddot{F}}{F}=0  \label{88}
\end{equation}
and $E_{0}-E_{2}=0$ gives
\begin{equation}
-\frac{\ddot{A}}{A}-\frac{\ddot{B}}{B}+\frac{(\dot{B})^{2}}{B^{2}}+\frac{
\dot{A}\dot{B}}{AB}-\frac{A^{2}}{2B^{4}}+\frac{1}{B^{2}}+\frac{\dot{B}\dot{F}
}{BF}-\frac{\ddot{F}}{F}=0.  \label{89}
\end{equation}
\textbf{Constant Curvature Solution:}

Utilizing the constant curvature condition (\ref{21}), the
Eqs.(\ref{88}) and (\ref{89}) take the form
\begin{eqnarray}
\frac{-2\ddot{B}}{B}+\frac{2\dot{A}\dot{B}}{AB}+\frac{A^{2}}{2B^{4}}=0,
\label{90}\\
-\frac{\ddot{A}}{A}-\frac{\ddot{B}}{B}+\frac{(\dot{B})^{2}}{B^{2}}+\frac{%
\dot{A}\dot{B}}{AB}+\frac{1}{B^{2}}-\frac{A^{2}}{2B^{4}}=0.
\label{91}
\end{eqnarray}
To solve these equations, we put $A(t)=2B(t)$ so that
Eqs.(\ref{90}) and (\ref{91}) reduce to
\begin{equation}
-\frac{\ddot{B}}{B}+\frac{(\dot{B})^{2}}{B^{2}}+\frac{1}{B^{2}}=0,
\label{92}
\end{equation}%
\begin{equation}
-\frac{\ddot{B}}{B}+\frac{(\dot{B})^{2}}{B^{2}}-\frac{1}{B^{2}}=0.
\label{93}
\end{equation}%
After addition the last two equations yield the following single
equation
\begin{equation}
-\frac{\ddot{B}}{B}+\frac{(\dot{B})^{2}}{B^{2}}=0,  \label{94}
\end{equation}
which can be easily solved by making the assumption,
$B(t)=e^{\lambda (t)}$ and yields
\[
\ddot{\lambda}=0.
\]
The solution of the last equation is obvious that $\lambda (t)$
must be a linear function of $t$, i.e. ,$ \lambda (t)=ft+g$, where
$f$ and $g$ are any arbitrary constants. Consequently, we obtain
the solution given by
\begin{eqnarray}
ds^{2}&=&dt^{2}-4e^{2(f\ t\ +\ g)}dx^{2}-e^{2(f\ t\ +\
g)}\{dy^{2}+ (4\cos ^{2}y+\sin ^{2}y)
dz^{2}\}\nonumber\\
&+&8\cos ye^{2(f\ t\ +\ g)}dxdz. \label{95}
\end{eqnarray}
In this case, the corresponding Ricci scalar turns out to be
constant, i.e., $R=-12f^{2}$.

\subsubsection{Case $~\gamma$:}

For this case, the Metric-III takes the following form
\begin{eqnarray}
ds^{2}&=&dt^{2}-A^{2}(t)dx^{2}-B^{2}(t)dy^{2}-\left\{
A^{2}(t)\cosh
^{2}y+B^{2}(t)\sinh ^{2}y\right\} dz^{2}\nonumber \\
&+&2\sinh yA^{2}(t)dxdz. \label{96}
\end{eqnarray}
The corresponding Ricci scalar has been evaluated, using the
components of Ricci tensor, as
\begin{equation}
R=\frac{-1}{2AB^{4}}[4\ddot{A}B^{4}+8\ddot{B}AB^{3}+8\dot{A}\dot{B}B^{3}+4A(
\dot{B})^{2}B^{2}-4AB^{2}-A^{3}].  \label{97}
\end{equation}
From Eq.(\ref{18}), $E_{0}-E_{1}=0$ yields
\begin{equation}
-\frac{2\ddot{B}}{B}+\frac{2\dot{A}\dot{B}}{AB}+\frac{A^{2}}{2B^{4}}+\frac{
\dot{A}\dot{F}}{AF}-\frac{\ddot{F}}{F}=0;  \label{98}
\end{equation}%
$E_{0}-E_{2}=0$ yields
\begin{equation}
-\frac{\ddot{A}}{A}-\frac{\ddot{B}}{B}+\frac{(\dot{B})^{2}}{B^{2}}+\frac{
\dot{A}\dot{B}}{AB}-\frac{A^{2}}{2B^{4}}-\frac{1}{B^{2}}+\frac{\dot{B}\dot{F}
}{BF}-\frac{\ddot{F}}{F}=0 \label{99}
\end{equation}%
and $E_{1}-E_{2}=0$ implies that
\begin{equation}
-\frac{\ddot{A}}{A}+\frac{\ddot{B}}{B}-\frac{\dot{A}\dot{B}}{AB}+\frac{(\dot{
B})^{2}}{B^{2}}-\frac{A^{2}}{B^{4}}-\frac{1}{B^{2}}+\frac{\dot{B}\dot{F}}{BF}
-\frac{\dot{A}\dot{F}}{AF}=0. \label{100}
\end{equation}\\
\textbf{Constant Curvature Solution: }

By making use of the constant curvature condition (\ref{21}), the
Eqs.(\ref{98})-(\ref{100}) reduce to
\begin{eqnarray}
\frac{-2\ddot{B}}{B}+\frac{2\dot{A}\dot{B}}{AB}+\frac{A^{2}}{2B^{4}}=0.
\label{101}
\\
-\frac{\ddot{A}}{A}-\frac{\ddot{B}}{B}+\frac{(\dot{B})^{2}}{B^{2}}+\frac{%
\dot{A}\dot{B}}{AB}-\frac{1}{B^{2}}-\frac{A^{2}}{2B^{4}}=0.
\label{102}
\\
-\frac{\ddot{A}}{A}+\frac{\ddot{B}}{B}-\frac{\dot{A}\dot{B}}{AB}+\frac{(\dot{%
B})^2}{B^{2}}-\frac{1}{B^{2}}-\frac{A^{2}}{B^{4}}=0.  \label{103}
\end{eqnarray}
Subtracting Eq.(\ref{101}) from Eq.(\ref{102}) we obtain the
Eq.(\ref{103}), which yields a trivial constant curvature solution
for $A=a$  and $B=b$ and $b=\pm a$. Consequently, the solution is
given by
\begin{eqnarray}
ds^{2}&=&dt^{2}-a^{2}dx^{2}-a^{2}dy^{2}-\left\{ a^{2}\cosh
^{2}y+a^{2}\sinh ^{2}y\right\} dz^{2}\nonumber \\
&+&2\sinh ya^{2}dxdz. \label{104}
\end{eqnarray}

\section{Summary and Discussion}

The study of $f(R)$ models has been carried out by many
researchers during the last two decades. They discussed
classification and comparison of the different approaches of
$f(R)$ theories of gravity. Also, different f(R) models have been
introduced to evaluate the energy density in $f(R)$ theory of
gravity. In this paper, we have solved the field equations of
$f(R)$ theory of gravity for LRS spacetimes by using metric $f(R)$
gravity approach.

As the field equations of $f(R)$ theory of gravity are highly
non-linear and complicated to be solved analytically due to the
arbitrary function $F$ so we use the assumption of constant scalar
curvature. This assumption is found to be the most appropriate and
make the field equations solvable in some cases. However, it is
not always be assured that the constant scalar curvature
assumption yield a solution.

The whole family of LRS spacetimes is represented by three
metrics, i.e., Metric-I, Metric-II and Metric-III.  The Metric-I
yields further three cases. The \texttt{Case(a)} yields four
non-trivial solutions as given by Eqs.(\ref{26}), (\ref{27}),
(\ref{36}) and (\ref{42}). While the \texttt{Case(b)} and
\texttt{Case(c)} give the nontrivial solutions given by
Eqs.(\ref{53}) and (\ref{62}) respectively. The Metric-II yields
only one non-trivial solution given by Eq.(\ref{73}). There arise
three cases for the Metric-III, i.e., Case$(\alpha)$,
Case$(\beta)$ and the Case$(\gamma)$. We obtain only one
nontrivial solution for the each Cases $(\alpha)$ and $(\beta)$
given by Eqs.(\ref{85}) and (\ref{95}) respectively. While the
Case$(\gamma)$ has been solved for trivial case only as given by
Eq.(\ref{104}).

The above mentioned solutions of LRS space-times in $f(R)$ theory
of gravity are given in the form of tables below:

\begin{center}
Table 1. Solution of Metric-I
\end{center}
\begin{center}
\begin{tabular}{|l|l|}
\hline Case & ~~~~~~~~~~~~~~~~~~~Solution \\ \hline Case aI(i) &
$ds^{2}=dt^{2}-(K_{1})^{2}t^{-\frac{2}{3}}dx^{2}-(K_{2})^{2}t^{
\frac{4}{3}}(dy^{2}+y^{2}dz^{2}).$ \\ \hline
Case aI(ii) & $%
ds^{2}=dt^{2}-(K_{1})^{2}t^{2}dx^{2}-(K_{2})^{2}(dy^{2}+y^{2}dz^{2}).$ \\
\hline Case aII(i) & $
ds^{2}=dt^{2}-(b^{2}t-b^{2}c)^{2}dx^{2}-(e)^{4a}(dy^{2}+y^{2}dz^{2}).$ \\
\hline
Case aII(ii) & $ds^{2}=dt^{2}-e^{4(ft+\bar{g}%
)}dx^{2}-e^{4(ft+g)}[dy^{2}+y^{2}dz^{2}].$ \\ \hline Case b &
$ds^{2}=dt^{2}-(K_{1}t)^{2}dx^{2}-(K_{2})^{2}[dy^{2}+\sin
^{2}ydz^{2}]$ \\ \hline Case c &
$ds^{2}=dt^{2}-(K_{1}t)^{2}dx^{2}-(K_{2})^{2}[dy^{2}+\sinh
^{2}ydz^{2}].$ \\ \hline
\end{tabular}
\end{center}
\begin{center}
\bigskip Table 2. Solution of Metric-II
\end{center}
\begin{center}
\begin{tabular}{|l|l|}
\hline Case & ~~~~~~~~~~~~~~~Solution \\ \hline 1 &
$ds^{2}=dt^{2}-e^{4a}dx^{2}-e^{4(f\ t\ +\ g)}\ e^{2\
x}(dy^{2}+dz^{2}).$
\\ \hline
\end{tabular}
\end{center}
\begin{center}
Table 3. Solution of Metric-III
\end{center}

\begin{center}
\begin{tabular}{|l|l|}
\hline Case & ~~~~~~~~~~~~~~~~~~~~Solution \\ \hline Case $\alpha
$ & $ds^{2}=dt^{2}-e^{4(f\ t\ +\ g)}(e^{4(f\ t\ +\
g)}dx^{2}-dy^{2}+2e^{4(f t+ g)}dxdz$\\&$~~~~~~~~~~~~~~~~-\left\{ e^{4(f\ t\ +\ g)}\frac{y^{4}
}{4}+y^{2}\right\} dz^{2}).$
\\ \hline Case $\beta $ & $ds^{2}=dt^{2}-e^{2(f\ t\ +\
g)}(4dx^{2}-dy^{2}-\left\{ 4\cos ^{2}y+\sin ^{2}y\right\}
dz^{2}$\\
&$~~~~~~~~~~~~~~~~+8\cos ydxdz.$ \\ \hline Case $\gamma $ &
$ds^{2}=dt^{2}-a^{2}dx^{2}-a^{2}dy^{2}-\left\{ a^{2}\cosh
^{2}y+a^{2}\sinh^{2}y\right\} dz^{2}$\\
&$~~~~~~~~~~~~~~~+2\sinh ya^{2}dxdz.$ \\
\hline
\end{tabular}
\end{center}

\vspace{0.25cm}

For the solution given in case aI(i), case aI(ii) and case aII(ii)
corresponds to the solution given in table $1$ and the Eq.(39) of
the \cite{56} respectively. It is mentioned here that the trivial
solution of case b, case c and the metric-II corresponds to the
solution given in Eq.(32), Eq.(33) and Eq.(50) of the \cite{56}
respectively.

\end{document}